\documentclass[pre,aps,twocolumn,showpacs,floatfix]{revtex4}
\usepackage{graphics}
\usepackage{epsfig}
\usepackage{amsmath}

\begin{document}

\title{Surface pattern formation and scaling described by conserved 
lattice gases}

\author{G\'eza \'Odor (1), Bartosz Liedke (2) and Karl-Heinz Heinig (2)}

\affiliation{(1) Research Institute for Technical Physics and
  Materials Science, \\
P.O.Box 49, H-1525 Budapest, Hungary\\
(2) Institute of Ion Beam Physics and Materials Research \\
Forschungszentrum Dresden - Rossendorf \\
P.O.Box 51 01 19, 01314 Dresden, Germany}    

\begin{abstract}
We extend our $2+1$ dimensional discrete growth model
(PRE 79, 021125 (2009)) with conserved, local exchange dynamics 
of octahedra, describing surface diffusion. 
A roughening process was realized by uphill diffusion and curvature dependence.
By mapping the slopes onto particles two-dimensional, nonequilibrium 
binary lattice model emerge, in which the (smoothing/roughening) surface 
diffusion can be described by attracting or repelling motion of 
oriented dimers. The binary representation allows simulations
on very large size and time scales.
We provide numerical evidence for Mullins-Herring or molecular 
beam epitaxy class scaling of the surface width.
The competition of inverse Mullins-Herring diffusion with 
a smoothing deposition, which corresponds to a Kardar-Parisi-Zhang
(KPZ) process generates different patterns: dots or ripples.
We analyze numerically the scaling and wavelength growth behavior 
in these models. In particular we confirm by large size 
simulations that the KPZ type of scaling is stable against the addition of
this surface diffusion, hence this is the asymptotic behavior of the 
Kuramoto-Sivashinsky equation as conjectured by field theory in two
dimensions, but has been debated numerically.
If very strong, normal surface diffusion is added to a KPZ process
we observe smooth surfaces with logarithmic growth, which can
describe the mean-field behavior of the strong-coupling KPZ class. 
We show that ripple coarsening occurs if parallel surface currents
are present, otherwise logarithmic behavior emerges.
\end{abstract}
\maketitle

\section{Introduction}

In nanotechnologies large areas of nanopatterns are needed, which can be
fabricated today only by expensive techniques, e.g. electron beam 
lithography or direct writing with electron and ion beams.
Besides the conventional 'top-down' technologies, which use masks,
photoresits, ... etc. to create structures on the surfaces, nowadays
'bottom-up' approaches are getting close to achieve the same results
more efficiently. In that case the self-assembly of patterns of
large areas is facilitated in a cost effective way \cite{FacskoS}.
This has led reopening of the research for fundamental theoretical 
understanding of the ion-beam induced surface patterning and scaling
\cite{MCB02}, which was flourishing at the end of the previous century 
\cite{HZ95,barabasi}. Although the basic universality classes and
important models have been explored, many notoriously difficult fundamental
questions have been unanswered. Perturbative renormalization group methods 
and analytical tools have limited applicability and precise numerical 
simulations, approaching asymptotic scaling regimes were feasible 
in one dimension mainly.

One of the most fundamental problem of kinetic roughening can
be characterized by the Kardar-Parisi-Zhang (KPZ) equation \cite{KPZeq}. 
The KPZ has been found to describe other important physical phenomena 
such as randomly stirred fluid \cite{forster77}, dissipative transport 
\cite{beijeren85,janssen86}, directed polymers \cite{kardar87} and the 
magnetic flux lines in superconductors \cite{hwa92}.
Therefore, we started our studies, by setting up the simplest
possible microscopic model exhibiting this behavior 
\cite{asep2dcikk,asepddcikk}.

The KPZ is a non-linear stochastic differential equation, describing
the dynamics of growth processes in the thermodynamic limit specified
by the height function $h({\bf x},t)$
\begin{equation}
\label{KPZ-e}
\partial_t h({\bf x},t) = v + \sigma\nabla^2 h({\bf x},t) + 
\lambda_2(\nabla h({\bf x},t))^2 + \eta({\bf x},t) \ .
\end{equation}
Here $v$ and $\lambda_2$ are the amplitudes of the mean and local growth 
velocities respectively, $\sigma$ is a smoothing surface tension coefficient
and $\eta$ roughens the surface by a zero-average, Gaussian noise field 
exhibiting the variance
\begin{equation}
\langle\eta({\bf x},t)\eta({\bf x'},t')\rangle 
= 2 D \delta^d ({\bf x-x'})(t-t') \ .
\end{equation}
We denote the spatial dimensions of the surface by $d$ and the 
noise amplitude by $D$.
The pure KPZ equation is exactly solvable in $1+1d$  \cite{kardar87},
but in higher dimensions only approximate solutions are available
(see \cite{barabasi}). In $d>1$ spatial dimensions due to the
competition between the roughening and smoothing, models 
characterized by (\ref{KPZ-e}), exhibit a roughening phase transition 
between a weak-coupling regime ($\lambda_2<\lambda^*_2$), governed by 
the $\lambda_2=0$ Edwards-Wilkinson (EW) fixed point \cite{EWc}, 
and a strong coupling phase.
The strong coupling fixed point is inaccessible by perturbative 
renormalization method. Therefore, the KPZ phase space has been
the subject of controversies and the value of the upper critical 
dimension is an active field of studies for a long time.

Mapping of surface growth onto reaction-diffusion system allows effective 
numerical simulations and better understanding of basic universality 
classes \cite{dimerlcikk,Orev,Obook08}. The principal aim of this paper 
is to show that some of most fundamental growth processes can be well 
described by the simplest, restricted solid on solid (RSOS) model with 
$\Delta h=\pm 1$. This strong condition enables a mapping onto binary 
lattice gases and facilitates to create fast algorithms.
We will discuss models, which follow universal scaling laws and exhibit 
pattern formation.
Although the understanding of coarsening phenomena \cite{B94} of 
surface patterns has been developing rapidly by continuum approaches 
\cite{CMC09} and the agreement with the ion beam induced nanopattern 
experiments is improving \cite{KCFM09}, the identification 
of the various coarsening scenarios is still a theoretical challenge
\cite{GGVGC10}.
Our approach is based on the KPZ models we introduced very 
recently \cite{asep2dcikk,asepddcikk}, hence for the sake of 
completeness we review them now.

In one dimension a discrete RSOS realization of the KPZ growth is 
equivalent \cite{kpz-asepmap,meakin} to the asymmetric simple 
exclusion process (ASEP) of particles \cite{Ligget}, while we have 
shown that this 'roof-top model' can be generalized to higher 
dimensions \cite{asep2dcikk,asepddcikk}.
This mapping is interesting not conceptually only, linking 
nonequilibrium surface growth with the dynamics of driven 
lattice gases \cite{S-Z,DickMar}, but provides an efficient 
numerical simulation tool for investigating debated and 
unresolved problems. 

\begin{figure}[ht]
\begin{center}
\epsfxsize=70mm
\epsffile{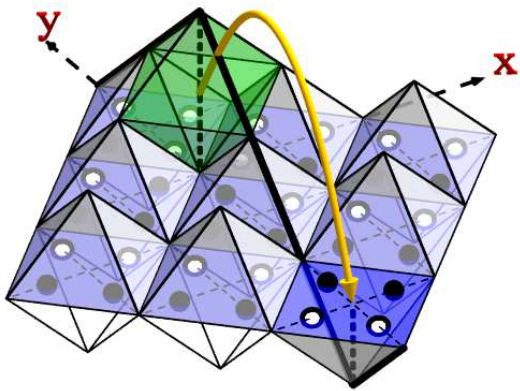}
\caption{(Color online) Mapping of the $2+1$ dimensional surface
diffusion onto a $2d$ particle model (bullets).
Surface attachment (with probability $p$) and detachment 
(with probability $q$) corresponds to Kawasaki exchanges of particles, 
or to anisotropic migration of dimers in the bisectrix direction
of the $x$ and $y$ axes. The crossing points of dashed lines
show the base sub-lattice to be updated. Thick solid line on 
the surface shows the $y$ cross-section, reminding us to the 
one-dimensional roof-top model. When the shown desorption/absorption 
steps are executed simultaneously, they realize a surface diffusion 
step of size $s=3$ along the $y$ axis. This corresponds to a pair of
dimer movement, a repulsion in case of this smoothing reaction.}
\label{2dM}
\end{center}
\end{figure}

The surface built up from the octahedra can be represented by the
edges meeting in the up/down middle vertexes (see Fig.~\ref{2dM}).
The up edges, in $\chi=x$ or $\chi=y$ directions at the lattice site
$i,j$ are represented by $\sigma_{\chi}(i,j)=+1$, while the down ones 
by $\sigma_{\chi}(i,j)=-1$ slopes. Therefore, we approximate
surfaces using RSOS model with $\Delta h=\pm 1$.

In this paper we show that one can describe various, more complex surface 
process without the need of having larger $\Delta h$ height differences.
These are built up from the basic octahedron deposition/removal processes
\cite{asep2dcikk}.
Let us remind the reader that a single site deposition flips four edges,
which means two '$+1$'$\leftrightarrow$'$-1$' (Kawasaki) exchanges:
one in the $x$ and one in the $y$ direction. This can be described by the 
generalized Kawasaki update
\begin{equation}
\left(
\begin{array}{cc}
   -1 & 1 \\
   -1 & 1
\end{array}
\right)\overset{p}{\underset{q}{\rightleftharpoons }}\left(
\begin{array}{cc}
   1 & -1 \\
   1 & -1
\end{array}
\right) \ , \label{rule}
\end{equation}
with probability $p$ for attachment and probability $q$ for detachment.

We can also call the '$+1$'-s as particles and the '$-1$'-s as holes 
living on the base square lattice, thus an attachment/detachment update
corresponds to a single step motion of an oriented dimer in the
bisectrix direction of the $x$ and $y$ axes. We update the 
neighborhood of the sub-lattice points, which are the 
crossing-points of the dashed lines.
In \cite{asep2dcikk,asepddcikk} we derived how this mapping 
connects the microscopic model to the KPZ equation and investigated 
the surface scaling numerically.
Our best estimates obtained by simulations up to sizes $L=2^{15}$
for the two dimensional KPZ universality class is in agreement 
with the operator product expansion result \cite{L98}
\begin{equation} \label{2dkpzexps}
\alpha=0.395(5), \ \ \beta=0.245(5), \ \ z=1.58(10) 
\end{equation}
within error margin \cite{asepddcikk}.

Besides, presenting some more interesting results about the universal
scaling behavior of KPZ in this paper we move further, and extend our 
mapping for describing more complex surface reactions. In particular,
we investigate models with surface diffusion processes, which are relevant 
in material science. We show the emergence of patterns and      
follow their coarsening dynamics.

\subsection{The simulations}

Although the bit-coded simulations are run on the underlying 
conserved lattice gas of size $L\times L$, starting from 
$h_{1,1}=1$ we reconstruct the surface heights from the differences
\begin{equation}
h_{i,j} = \sum_{l=1}^i \sigma_x(l,1) + \sum_{k=1}^j \sigma_y(i,k)
\end{equation}
at certain sampling times ($t$), selected with power-law
increasing time steps and calculate its width:
\begin{equation}
W(L,t) =
\Bigl[
\frac{1}{L^2} \, \sum_{i,j}^L \,h^2_{i,j}(t)  -
\Bigl(\frac{1}{L} \, \sum_{i,j}^L \,h_{i,j}(t) \Bigr)^2
\Bigr]^{1/2} \ .
\end{equation}
In the absence of any characteristic length, the surface is expected
to follow Family-Vicsek scaling~\cite{family}, when we start from a flat
initial condition
\begin{eqnarray}
\label{FV-forf}
W(L,t) &\propto& t^{\beta} , \ \ {\rm for} \ \ t_0 << t << t_s \\
       &\propto& L^{\alpha} , \ \ {\rm for} \ \ t >> t_s \ . \label{FV-a}
\end{eqnarray}
Here, $\alpha$ is the roughness exponent for $t>>t_s$ when the correlation 
length has exceeded the linear system size $L$;
and $\beta$ is the surface growth exponent, which describes the
time evolution for earlier (non-microscopic $t>>t_0$) times.
The dynamical exponent $z$ can be expressed by the ratio
\begin{equation}\label{zlaw}
z = \alpha/\beta \ .
\end{equation}
However in case of pattern formation multi-scaling is present in the
system and the roughness exponent calculated in different window sizes is not
constant and satisfies a different, anomalous scaling law 
(see \cite{lopez96,krug-rev}).

The morphology of pattern formation in experiments is usually followed by
the measurement of some characteristic size, e.g. the wavelength in 
periodic structures, which evolves non-trivially with time. We can
easily define and measure such quantity in our model.
By following the up, or down slopes in the $x$ direction we have 
strings of '$1$'-s or '$-1$' of length $s_k$ in the $k$-th slice 
of the lattice gas in the $y$ direction.
We shall characterize patterns by calculating the ($y$) average of the 
longest $s_k$ value
\begin{equation}\label{lambda}
\lambda = 1/L \sum_{k=1}^L \max(s_k)  \ .
\end{equation} 
This characteristic length, corresponds to the longest $x$ slope, or to
the slowest $x$ mode in the Fourier decomposition, will provide information 
about scaling of the wavelength in our analysis.

By the simulations we apply periodic boundary conditions in both directions,
and start from the flat space corresponding to a zig-zag configuration of the
slopes (see Fig.\ref{2dM}), therefore it has a small initial width 
$W^2(L,0)=1/4$. Before the scaling analysis we always subtract this 
constant, being the leading-order correction, from the raw data. 
Averaging was done usually for $100-200$ samples for each parameter value.

In practice each lattice site can be characterized by the 16 different
local slope configurations, but we update it only when the condition
(\ref{rule}) is satisfied. Furthermore due to the surface continuity not all 
configurations may occur and we can describe a lattice site by using
only 2 bits. This permits efficient storage management in the 
memory of the computer and large system sizes. The updates can be 
performed by logical operations either on multiple samples at 
once or on multiple (not overlapping) sites at once. Our bit-coded
algorithm proved to be $\sim 40$ times faster than the conventional
FORTRAN 90 code. A crucial point is to use a good, high resolution 
random number generator, because in case of the $p=1$ KPZ process the only
source of randomness is the site selection, which must be done in
a completely uniform way. Otherwise we realize a KPZ with quenched disorder
which belongs to a different universality class (see \cite{Obook08}).
We used the latest Mersenne-Twister generator \cite{MT} in general, 
which has very good statistical properties and which is very fast, especially
by the SSE2 instructions. But we tested our results using other 
random number generators as well.
 
An elementary Monte Carlo step (MCs) starts with a random site selection.
This is followed by testing if the place is appropriate for update
i.e. 'roof-top' for detachments or 'valley-bottom' for attachment 
(\ref{rule}). The update is done with the prescribed $p$ and $q$ 
probabilities and the time is incremented by $1/L^2$, such that one 
MCs corresponds to a full lattice update. Throughout the paper we use 
this unit of time.

\subsection{Generalizations of the octahedron model}

An obvious first step is to combine the deposition and the removal 
processes creating a conserved dynamics. A simultaneous octahedron
detachment and deposition in the neighborhood can realize an elementary 
diffusion step.
Surface diffusion is a much studied basic process \cite{krug-rev}.
Several atomistic models have been constructed and investigated with the
aim of realizing Mullins-Herring (MH) diffusion \cite{herring50,mullins} 
and scaling (for a recent review see \cite{MBEpers}). 

The Langevin equation of MH is a linear one, with a $\nabla^4$
lowest order gradient term
\begin{equation}
\label{MH0e}
\partial_t h({\bf x},t)  = \nu_4\nabla^4 h({\bf x},t)  + \eta({\bf x},t) \ .
\end{equation}
emerging as the result of a curvature driven surface current
$j({\bf x},t) \propto \nabla (\nabla^2 h({\bf x},t))$, which obeys 
the conservation law
\begin{equation}
\partial_t h({\bf x},t) + \nabla j({\bf x},t) = \eta({\bf x},t) \ .
\end{equation}
Here the noise $\eta({\bf x},t)$ is a non-conserved, Gaussian white one, 
which can be the result of fluctuations in the ion-beam intensity directed
against the surface.
This equation is exactly solvable, and exhibits a scaling invariance
of the roughness characterized by the exponents:
\begin{equation}\label{MHclass}
\alpha=(4-d)/2, \ \ \ \beta=(4-d)/8, \ \ \ z=4 
\end{equation}
below the upper critical dimension $d_c=4$.

Microscopic models realizing this behavior are mainly unrestricted 
solid on solid (SOS) type, that can provide steep slopes and strong 
curvatures necessary for the $\alpha\ge 1$ roughness exponent 
in one and two dimensions. However it turned out that the asymptotic 
universality class of the various limited mobility growth models 
is a surprisingly subtle issue. Many of the earlier findings proved 
to be incorrect due to pathologically slow crossover and extremely 
long transient effects \cite{SCT02}.
Anomalous scaling, by which the local and global behavior is different, 
has been found to be relevant in such 'super-rough' models, where 
large local slopes are present 
\cite{BPFS98,fracture2,das0,mario,lopez96,krug91}.
In fact, according to our knowledge, only the "larger curvature 
(or Kim-Das Sarma) model" \cite{KFS94,krug} and the "$n=2$ model" of 
\cite{SP94} exhibit MH universality class scaling asymptotically.

The scaling in other 'atomistic' models crosses over to behavior dictated 
by more relevant terms in the sense of renormalization group. This can be the
EW class if $\nabla^2(h)$ \cite{family86} is present, or the Molecular Beam
Epitaxy (MBE) class in case of the fourth order non-linearity 
$\nabla^2[\nabla(h)^2]$ \cite{WV90,DT91,KPK,das2} 
\footnote{Cubic non-linearity is also possible by the same symmetries, 
however not much realization  or occurrence has been found}. 
The nonlinear MBE equation:
\begin{equation}
\label{CKPZeq}
\partial_t h  = \nu_4\nabla^4 h + \lambda_{22}\nabla^2[\nabla(h)^2] + \eta 
\end{equation}
with non-conserved, Gaussian noise $\eta$ is just the conserved version 
of KPZ (CKPZ) and exhibits the following scaling exponents:
\begin{equation}
\alpha=(4-d)/3, \ \ \ \beta=(4-d)/(8+d), \ \ \ z=(8+d)/3  \ .
\end{equation}
As we can see all exponents are smaller than those of the linear MH 
(\ref{MHclass}) class values and differ from those of the two-dimensional
KPZ class (\ref{2dkpzexps}) significantly. 

Here we present RSOS models with $\Delta h=\pm 1$ height restriction 
within the framework of our previous approach \cite{asep2dcikk}, 
which in the limit of weak external noise exhibit MH or MBE scaling. 
Due to the simple construction these can be mapped onto
lattice gases of diffusing dimers, allowing an easier way to study 
the effects of MH and MBE sub-processes of more complex system.
Earlier less restrictive RSOS models were used to describe 
these classes especially in one dimension (for a review see
\cite{MBEpers}).

A further step in generalizing our surface models will be 
the combination of different sub-processes resulting in 
nonequilibrium system. For example, by adding a competing MH diffusion 
to the KPZ updates one can model the noisy Kuramoto-Sivashinsky 
(KS) equation
\cite{ks:1977,sivashinsky:1979}.
\begin{equation}
\label{KS-e}
\partial_t h = v + \sigma\nabla^2 h -\nu_4\nabla^4 h + 
\lambda_2(\nabla h)^2 + \eta \ .
\end{equation}
Here the surface tension coefficient $\sigma$ is negative (in contrast with
the KPZ), whereas $\nu_4$ is a positive surface diffusion coefficient.
However in our simulations we realize a gauge transformed situation: 
normal KPZ (with positive $\sigma$) competing with an inverse MH (iMH) with 
negative $\nu_4$ \footnote{The opposite, positive $\nu_4$ case corresponds to
normal, smoothing surface diffusion.}.
In KS even the deterministic variant ($\eta=0$) exhibits spatio-temporal chaos,
and is useful to describe pattern formation, such as chemical
turbulence and flame-front propagation \cite{ks:1977}.
There are other physical systems, including ion sputtering, where
the noisy version of the KS equation is used \cite{CL95}.
In one dimension, field theory has proved that KS belongs to
the $1+1$ dimensional KPZ universality class \cite{CL95}. 
However in higher dimensions perturbative field theory cannot 
access the strong coupling fixed point.
Numerical studies \cite{Proc92,JHP93,DZLW99} have provided controversial 
results in $2+1$ dimensions, hence it remained a controversial 
and challenging problem to clarify the asymptotic scaling behavior of KS
\cite{lvov-proc92,lvov-leb93,lvov-proc94,JHP94}.

It was pointed out \cite{RLP90,krug99} that grooved phases and growth
instabilities may emerge as the consequence of broken detailed balance 
condition:
\begin{equation}\label{dbal}
P(\{i\}) w_{i\to i'} = P(\{i'\}) w_{i'\to i}
\end{equation}
where $P(\{i'\})$ denotes the probability of the state $\{i\}$
and $w_{i\to i'}$ is the transition rate between states $\{i\}$ and
$\{i'\}$.
This means that complex structures and patterns can emerge in nonequilibrium
system.
In one dimension a model of massive particles exhibiting momentum 
has been shown to exhibit KS scaling behavior \cite{RK95}.
Later another one-dimensional RSOS growth model was constructed 
\cite{krug-hont},
in which deposition and diffusion of single ad-atoms were competing.
It was suggested that large-scale behaviour could be described by the 
noisy KS equation.

\section{Realizing the MBE surface diffusion}

As we have mentioned in the Introduction we generate surface diffusion as
the simultaneous adsorption/desorption of octahedra. Therefore, in
our model after an appropriate removal site (a roof-top) selection
has been done, we search for a valley bottom place in the 
neighborhood for deposition. The target site is chosen in the 
$\pm x$ or $\pm y$ direction, with the probabilities: 
$p_{+x}$, $p_{-x}$ or $p_{+y}$, $p_{-y}$ respectively 
(see Fig.\ref{2dM}). Throughout of our studies we normalized the
attempt probabilities. The maximal jump distance was fixed to be
$l_m\le 4$ lattice units following computer experiments. According to
the construction the nearest neighbour jump, corresponding to
intralayer diffusion, requires $l_m=2$, while larger jump sizes allow
intra-layer transport. To create MBE or MH behavior we must allow 
inter-layer transport. Considering jumps with $l_m \ge 3$ does not make 
difference in the scaling, larger jumps describe faster diffusion
on the expense of more CPU time. 
Having no apriori knowledge of the distance dependence of jump
sizes and expecting insensitivity of universal scaling on the rates of
short ranged interactions we used a constant probability for 
each direction (for more discussion see Sect.~\ref{IIA}).

To control this kind of surface diffusion we impose additional constraints
for the accepting a move. We have tried two kinds of rules based on
the local neighborhood configurations. The first one is very simple
and requires that the height of a particle at the final state, is higher
than that of its initial site
\begin{equation}\label{Rproc}
h_{fin}-h_{ini} \ge 0 \ ,
\end{equation} 
which makes a surface more rough (inverse/roughening diffusion).
The second one is based on the local curvature at the update sites as will 
be discussed in Sect.~\ref{IIB}.

\subsection{The larger height octahedron diffusion model} \label{IIA}

By this simple condition, we expect to generate rough surfaces of large 
curvatures, since an octahedron jumping to a higher position has 
usually smaller number of neighbours in the lateral direction (in 
this model we do not allow particles to evaporate from the bottom of 
a V shaped valley).
In the language of lattice gases regions of large curvature and maximal
slope correspond to large regions of dense particles and holes. 
An attractive interaction of the dimer moves enhances such segregation.  
Contrary, we can realize a smoothing/normal diffusion, when we accept 
jumps with the condition: $h_{fin}-h_{ini} \le 0$.

By increasing the local heights the formation of pyramid-like structures
,unstable growth, occurs similarly as in the $n=2$ SOS model \cite{SP94},
but in our case the slopes are limited to 45 degree. However, by appropriate
length rescaling any sharp, continuous surface can be approximated.   
In the forthcoming, we will investigate both the inverse and the normal
diffusion cases and call it the larger height octahedron diffusion 
(LHOD) model.
 
\begin{figure}[ht]
\begin{center}
\epsfxsize80mm
\epsffile{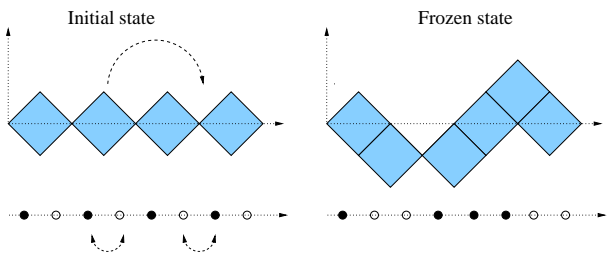}
\caption{(Color online) One dimensional view of a diffusion hop
of octahedra by 3 lattice units to the right. The slopes mapped 
onto the lattice gas shown below the surface. This roughening
surface diffusion move corresponds to two simultaneous, attracting 
Kawasaki exchanges of the gas (dimers in two dimensions). Starting
from the flat (zig-zag) initial state of the pure octahedron model 
the (\ref{Rproc}) process freezes following slopes of maximal 
length are developed, which can't be over-jumped.}
\label{freezing}
\end{center}
\end{figure}

Starting from a zig-zag initial condition, corresponding to the
flat surface, the hopping with the condition (\ref{Rproc}), one
can make the surface more rough, however after the slopes of size $l_m$ 
are developed the evolution stops, because the octahedra are 
not allowed to pass longer $45$ degree gradient sections 
(see Fig.~\ref{freezing}). 
To overcome this, we have been trying to allow arbitrarily long 
jump sizes (non-local model), with different, heuristic jump 
distance dependent probability functions.
We did not find an appropriate one that could produce the
expected MH scaling behavior, instead we realized L\'evy flight like
models with anomalous diffusion \cite{MK00} exhibiting non-universal scaling. 
In he light of recent field theoretical interest \cite{CuernoL} 
these can also be the target of further investigations.

However, non-local models are rather complex and connection to reality is
not always straightforward. Therefore, we followed an other strategy 
by adding a small amount of extra randomness of EW type 
to our short-range, binary RSOS model. 
This means the addition of random adsorption/removal events with
small probability $p=q<<1$ among the LHOD updates.
Such events can break up the barriers by splitting up the long monotonous slopes
built by the LHOD dynamics. If they are done very rarely, i.e. with less than 
$100$ times smaller probability than the diffusion attempts, they can influence 
the very late asymptotic scaling behavior only, causing an ultimate crossover 
to the EW class (see for example \cite{HW07}).

In reality, and in the models we are about to study, such randomizing effects 
are always present, thus we are satisfied, if we can confirm numerically the 
MBE scaling for intermediate times.
It was not our principal aim to provide a model, which exhibits pure 
MBE type of scaling in the thermodynamic limit.  
Therefore we have performed computer experiments on these models
as discussed below.
\begin{figure}[ht]
\begin{center}
\epsfxsize70mm
\epsffile{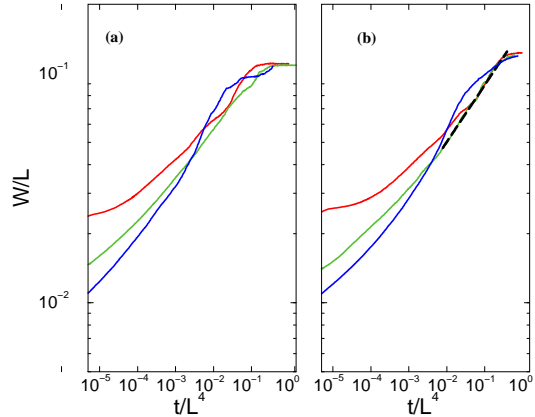}
\caption{(Color online) Data collapse of the anisotropic LHOD model assuming
MH class exponents for $p_{+x}=1$ (diffusion to the right)
with small EW noise $p=q=0.01$ (a), $p=q=0.005$ (b), for sizes $L=32,64,128$
(top to bottom at the left side). For the growth exponent fitting
(dashed line) results in $\beta=0.26(1)$.}
\label{mbe5}
\end{center}
\end{figure}

We found that the addition of the small (EW) noise ($p=q<<1$) sustains the
surface currents, and in case of spatially anisotropic diffusion ($p_{+x}=1$, 
$p_{-x}=p_{y}=p_{-y}=0$) the scaling becomes Mullins-Herring type 
(see Fig.~\ref{mbe5}(a)), characterized by the exponents 
$\alpha=1$, $\beta=1/4$, $z=4$ in two dimensions.
The collapse of curves is very good in the vertical direction, corresponding to
$\alpha=1$, and the horizontal scaling improves as we decrease the 
noise (see Fig.~\ref{mbe5}(b)). In opposite, by increasing the aplitude of the
EW noise, we can find better collapse with smaller dynamical exponent, 
describing the cross-over to the EW behavior, which has $z=2$.
The time dependence shows deviations from the pure scaling-law, especially 
for early times, still before the saturation the growth of $W(t)$ 
can be fitted with the exponent $\beta=0.26(1)$.
\footnote{Note however that for $\alpha\ge 1$, anomalous scaling also
occurs, which we don't study here.}
This suggests that in the $p=q\to 0$ limit the true MH scaling emerges.
Simulating larger sizes is very hard, because due to the large dynamical
exponent $z$ the saturation is shifted to very late times (for $L=128$ 
this happens for $t > 2\times 10^8$ MCs only).

\begin{figure}[ht]
\begin{center}
\epsfxsize80mm
\epsffile{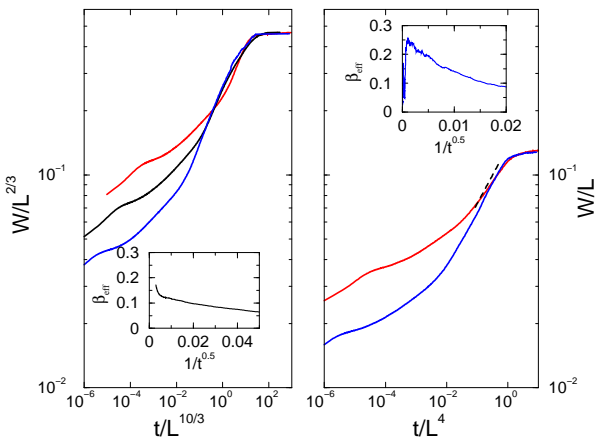}
\caption{(Color online) Data collapse of the isotropic LHOD model 
assuming MBE (left, $l_m=4$) and MH (right, $l_m=3$) class exponents 
in the presence of small EW noise $p=q=0.005$ in sizes 
$L=32,64,128$ (top to bottom at the left side). For the growth 
exponent power-law fitting (dashed line) results in $\beta=0.26(1)$. 
The inserts show the effective $\beta$ exponents for $L=64$.}
\label{mbei}
\end{center}
\end{figure}

In case of isotropic diffusion the scaling is MBE class type
in general (see Fig.~\ref{mbei}(a) for $l_m=4$), characterized by the 
exponents $\alpha=2/3$, $\beta=0.2$, $z=10/3$ in two dimensions. 
Therefore, the algorithm with the LHOD update (\ref{Rproc}) breaks 
the detailed balance condition (\ref{dbal}) and introduces a
non-linearity. 
However, this non-linearity is small and by decreasing $l_m$ 
it becomes even smaller (see Fig.~\ref{mbei} (b)). One can find a 
rather good collapse with the MH exponents in case of $l_m=3$.
We estimated the growth exponent by calculating the local slopes
\begin{equation}  \label{beff}
\beta_{eff}(t) = \frac {\ln W(t,L\to\infty) - \ln W(t^{\prime},L\to\infty)} 
{\ln(t) - \ln(t^{\prime})} \ .
\end{equation}
As one can read-off from the inserts of Fig.~\ref{mbei}, for $l_m=4$
this effective exponent extrapolates to the MBE value ($\beta=0.20(2)$), 
while for $l_m=3$  to $\beta=0.26(1)$ agreeing with the MH exponent in the 
$t\to\infty$ limit.
This means that the LHOD rule introduces more possibilities 
for breaking the detailed balance condition (\ref{dbal}), 
when more degrees of freedom (more directions or larger 
jumps sizes) are allowed.
This provides an explanation for the difference between the 
scaling behavior of the isotropic and anisotropic diffusion.
To avoid such non-linearity completely we introduced a more complex 
update rule, based on the local curvature conditions 
keeping the spatial symmetries.

\subsection{The larger curvature octahedron diffusion model} \label{IIB}

\begin{figure}[ht]
\epsfxsize60mm
\epsffile{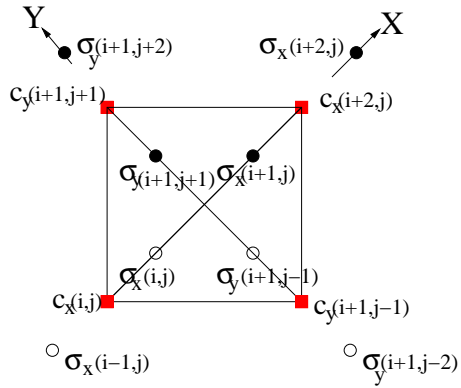}
\caption{Local slope $\sigma_{\chi}(i,j)$ (circles) and curvature 
$c_{\chi}(i,j)$ (squares) variables at an update site. 
Filled circles correspond to upward, empty ones to downward 
slopes of the surface. This plaquette configuration
models a valley bottom site, with the total curvature $H=4$.}
\label{MHupdate}
\end{figure}

In this section we describe a transition probability in addition to
the octahedron surface hopping model, which satisfies the detailed 
balance condition (\ref{dbal}), thus enables one to realize linear, 
equilibrium MH diffusion steps. 
The local curvature of the surface is calculated at the $4$ edges of 
squares of the projected octahedra. 
As Figure~\ref{MHupdate} shows one can describe the curvatures 
$c_{\chi}(i,j)$ ($\chi\in(x,y)$) by the products (or differences)
of the local slopes 
\begin{equation}
c_{\chi}(i,j) = \sigma_{\chi}(i,j)\sigma_{\chi}(i+1,j) \ .
\end{equation}
At each update we calculate the sum of the change of local curvatures
at the origin ($i,j$) and the target ($i',j'$) sites
\begin{equation}
\Delta H=\Delta\sum_{\chi=x,y}\sum_{\langle i,j\rangle}c_{\chi}(i,j) 
+ \Delta\sum_{\chi=x,y}\sum_{\langle i',j'\rangle} c_{\chi}(i',j') \ ,
\end{equation}
where $\langle \rangle$ denotes the plaquette neighborhood sites 
as shown on Fig.~\ref{MHupdate}. This gives maximal value: 
$H=4$ for a local tops and bottoms and the minimal value: $H=-4$ for 
a locally flat (zig-zag) configuration.
Using this value we accept the update with an Arrhenius type of 
probability
\begin{equation}
w_{i\to i'} = 1/2 [1 - a \tanh(-\Delta H^2)] \ ,
\end{equation}
where $a$ is a constant. This form is very similar to what was used 
in case of the one-dimensional $n=2$ model \cite{SP94} and enhances
(suppresses) roughening moves if $a>0$ ($a<0$) respectively. 
In \cite{SP94} symmetry arguments was applied for the lowest order 
series expansion of $w_{i\to i'}$ of the model to prove a
connection with the MH equation. Now similar derivation can
be done by extending the model for dimer variables in two dimensions.
In the forthcoming we shall call this the larger curvature
octahedron diffusion (LCOD) model.

\begin{figure}[ht]
\begin{center}
\epsfxsize70mm
\epsffile{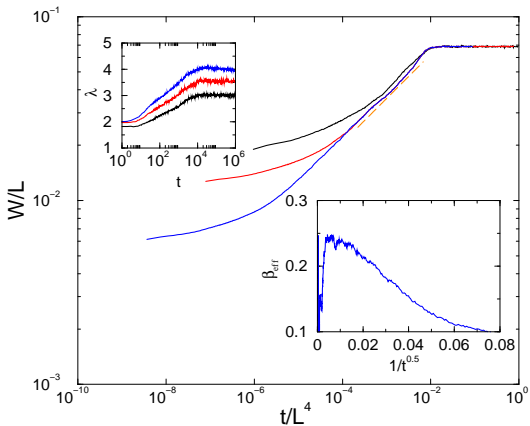}
\caption{(Color online) Scaling behavior of the isotropic LCOD model
for $L=32,64,128$ (top to bottom at the left side). The data collapse 
has been achieved with the MH class exponents. For the growth exponent fitting
(dashed line) results in $\beta=0.25(1)$. The insert on the right shows
the same by local slopes. The insert on the left shows the evolution of 
$\lambda$.}
\label{MHB}
\end{center}
\end{figure}

We have simulated the LCOD with the parameters: $a=0.1$, $l_m=3$, 
corresponding to inverse/roughening surface diffusion and $p=q=0.05$.
The scaling behavior has been found to agree very well with that of 
MH universality class values even in case of isotropic diffusion 
(see Figure~\ref{MHB}).
The effective $\beta_{eff}(t)$ converges to $\beta=0.25(1)$ 
before the saturation. The wavelength grows logarithmically in time
(see insert of Figure~\ref{MHB}) and after a steady state value is 
reached it scales logarithmically with the the system size too. 

For anisotropic diffusion ($p_{\pm x}=1, p_{\pm y}=0$) we have 
obtained similarly good MH class surface scaling, with logarithmic
time and size dependence of $\lambda$ again.

For $a=0$, without any EW noise, one can find logarithmic growth
in the LCOD model in time
\begin{equation}
W(t,L\to\infty) \propto \ln(t)
\end{equation}
and logarithmic surface roughness dependence of
\begin{equation}
W(t\to\infty,L) \propto \ln(L) \ .
\end{equation} 
Data collapse fitting for the dynamical exponent on the other 
hand results in space-time anisotropy with $z=4$.
This means that for the $a=0$, noiseless case we could realize
the universality class behavior of MH with conserved 
(purely diffusive) noise, characterized by the exponents:
$\alpha=\beta=0$, $z=4$ (see \cite{Obook08}).

\section{Pattern generation by competing inverse MH and KPZ processes}

As we have shown in the previous section, in the zero noise limit our 
RSOS surface diffusion processes generate growth with MBE or MH scaling 
behavior. In this section we investigate them in the presence of 
competing KPZ updates. In our simulations we initiate hopping
with probabilities: $p_{+x}$, $p_{-x}$ , $p_{+y}$, $p_{-y}$
alternately with the deposition (with probability $p$) and removal (with
probability $q$) processes. We follow the surface roughness and
pattern formation with the corresponding wavelength growth.

\subsection{Spatially anisotropic surface diffusion}

\begin{figure}[ht]
\begin{center}
\epsfxsize 90mm
\epsffile{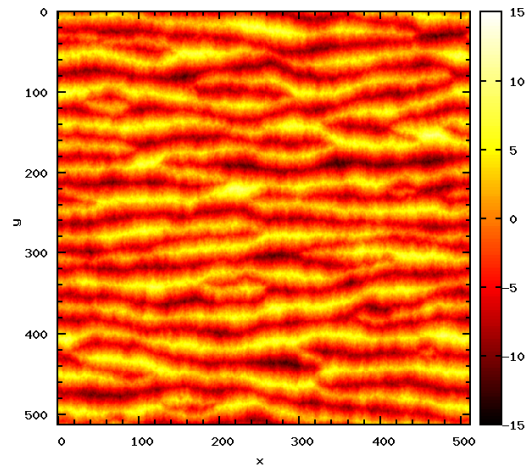}
\caption{(Color online) Snapshot of surface heights of the ripple 
patterns generated by the parameters $p_{\pm x}=1$, $p_{\pm y}=0$
(anisotropic, inverse MH) and $p=q=1$ adsorption/desorption
at $t=10^4$ MCs, in the LHOD model of linear size $L=512$. }
\label{AKS}
\end{center}
\end{figure}

From the point of pattern formation the LHOD and LCOD model behaves
differently. We always start the simulations from a flat surface
and watch if stable patterns can arise. 
In case of an anisotropic inverse LHOD model of diffusion probability 
$p_{\pm x}=1$, $p_{\pm y}=0$ a competing EW process always generates
ripple patterns as shown on Fig.~\ref{AKS}), which is stable for 
all $p=q\le 1$. This formation is metastable against KPZ (height anisotropy),
but for very large times (in the steady state) the ripples become
uneven, blurred and cut into smaller pieces.
The wavelength, defined as (\ref{lambda}) grows only a little 
(in a power-law manner) and saturates quickly. 

\begin{figure}[ht]
\begin{center}
\epsfxsize70mm
\epsffile{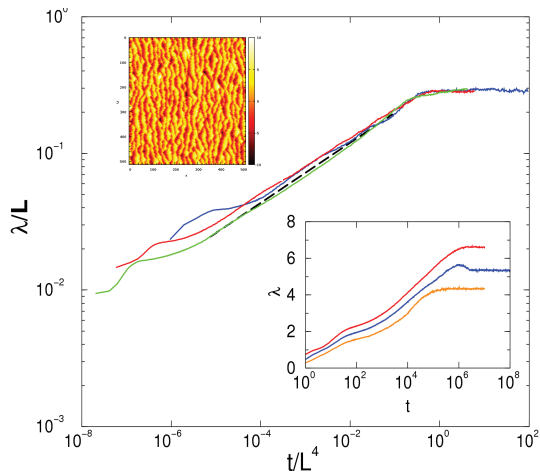}
\caption{(Color online) The wavelength growth in the LHOD model
for anisotropic diffusion with steady DC current $p_{+y}=1$, 
$p=q=0.005$ for sizes $L=32,64,128$ (top to bottom at the beginning). 
Dashed line: power-law fit with the exponent $\beta=0.24(1)$.
The left insert shows the corresponding pattern.
The right insert corresponds the isotropic diffusion case 
$p_{\pm x}=p_{\pm y}=1$, where $\lambda(t)$ grows logarithmically.}
\label{ALHODl}
\end{center}
\end{figure}

\begin{figure}[ht]
\begin{center}
\epsfxsize70mm
\epsffile{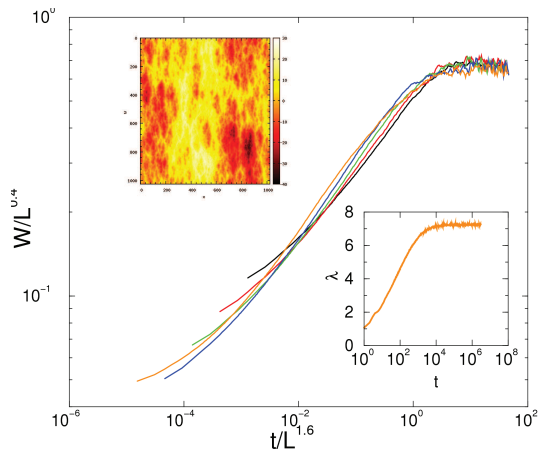}
\caption{(Color online) Data collapse for deposition ($p=1$, $q=0$)
and anisotropic, inverse diffusion $p_{\pm y}=1$ in the LHOD model
with KPZ class exponents for $L=64,128,256,512,1024$ 
(top to bottom curves at the right side). 
Right insert: $\lambda(r)$ for $L=1024$, left insert the blurred
ripple structure.}
\label{kpz+iad}
\end{center}
\end{figure}

However, if we create such anisotropy in which a steady, direct 
current (DC) flows, for example when only $p_{+x}=1$ and all the others 
are zero we can find a different behavior. In this case, for weak
EW or KPZ the ripples are not completely straight and exhibit a 
coarsening as
\begin{equation}
\lambda\propto t^{0.24(1)} \ .
\end{equation}
The ripples are more straight in the KPZ case than for up-down
isotropic deposition/removal.
Furthermore, a good data collapse can be obtained with the 
MH class exponents for sizes $L=32,64,128$ as shown on 
Fig.~\ref{ALHODl}.
However, this scaling can be destroyed by increasing the strength of
the non-conserved reaction and the power-law crosses over to logarithmic
growth of $\lambda$.
In case of strong KPZ deposition ($p=1$, $q=0$) the asymptotic 
scaling of the LHOD becomes completely KPZ type, the wavelength saturates 
very quickly (see Fig.~\ref{kpz+iad}) and the ripple structure smears.
In principle we should expect anisotropic KPZ behavior here, but 
it is well known that in two dimensions such spatial anisotropy
is irrelevant, hence the isotropic KPZ class behavior \cite{TF02}
is not surprising.
Note that our anisotropic KS model is different from what is called
and considered to be the "anisotropic KS" in the literature \cite{aniKS},
because in our case the surface diffusion (corresponding to the $\nabla^4$ term)
is anisotropic. Such models are very hard for analytic treatment and spatial
anisotropy is introduced in the $\nabla^2$ terms usually.

The anisotropic LCOD is less effective for ripple formation than the
LHOD. In this case the patterns are smoother, and the $\lambda$ scales 
logarithmically in time and by the size. The only exception is
when we allow a steady DC current again. In this case similar
power laws as shown on Fig.~\ref{kpz+iad} emerge for weak EW or KPZ.

When DC current is not allowed ($p_{\pm x}=1, p_{\pm y}=0$), only
spatial anisotropy in the LCOD and we add a KPZ ($p=1$, $q=0$) update 
we can see the emergence of KPZ scaling. In this case the wavelength 
depends logarithmically both on time and the size $L$. 
It is important to realize that for short, one decade length, 
time windows the wavelength growth can also be well fitted with a power-law
\begin{equation} \label{lamloggrow}
\lambda(t,L\to\infty) \propto t^{0.17(1)} \ ,
\end{equation}
which resembles to experimental results, but since the steady state values exhibit 
a clear logarithmic dependence on the sizes
\begin{equation} \label{lamlogrou}
\lambda(t\to\infty,L)  \propto \ln(L) \ ,
\end{equation}
we don't think that this 'power-law' fit would correspond to a real 
asymptotic behavior in the thermodynamic limit.  

\subsection{Spatially isotropic surface diffusion}

\begin{figure}[ht]
\begin{center}
\epsfxsize 90mm
\epsffile{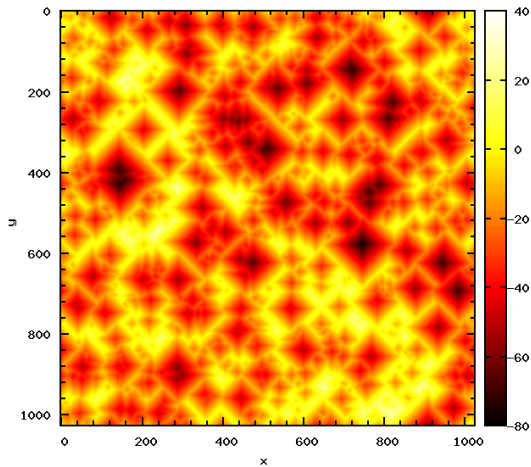}
\caption{(Color online) Snapshot of surface heights of the dot
patterns generated by the LHOD model with parameters: 
$p_{\pm x}=p_{\pm y}=1$ (isotropic, inverse MH) 
and $p=q=0.1$ at $t=10^4$ MCs.}
\label{KS}
\end{center}
\end{figure}

When the isotropic, inverse surface diffusion competes with the (smoothing)
EW process one can observe dot formation both in the LHOD and LCOD models. 
Fig.~\ref{KS} shows a snapshot of the growing dots for LHOD in the
presence of weak EW process. Here, one can see rectangular shaped 
patterns corresponding
to the lattice symmetry. In case of LCOD the contrast of the patterns is 
smoother, roughly circular.
However, this pattern coarsening is much slower than in case of ripples.

We shall discuss the LHOD results first, which according to our previous
numerical analysis, corresponds to the nonlinear equation (KPZ+MBE):
\begin{equation}
\label{NCKPZ}
\partial_t h  = \sigma\nabla^2 h + \lambda_2(\nabla h)^2 + \nu_4\nabla^4 h 
+ \lambda_{22}\nabla^2[\nabla(h)^2] + \eta
\end{equation}
In case of EW type of deposition/removal the nonlinear term vanishes
($\lambda_2=0$) and in fact we model the CKPZ behavior.
The characteristic size ($\lambda$) of the dots grows logarithmically
(see insert of Fig.~\ref{ALHODl}) in time and 
$\lambda_{max}\propto \ln(L)$.
The pattern formation is more pronounced in the LHOD case than 
in the LCOD model. We associate it to the up/down anisotropy present 
in the LHOD.

\begin{figure}[ht]
\begin{center}
\epsfxsize70mm
\epsffile{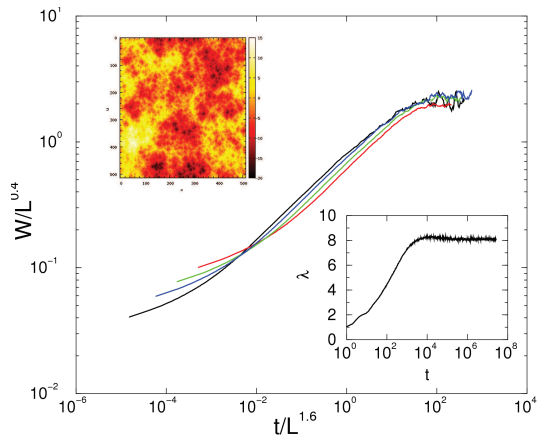}
\caption{(Color online) Data collapse of the $L=128,...1024$ LHOD model
($p_{\pm x}=p_{\pm y}=1$) with a competing deposition ($p=1$) process.
One can see a very slow crossover towards KPZ scaling.
The right insert shows the growth of $\lambda$ for $L=512$.
The left insert is a snapshot of the steady state, 
corresponding to the smeared KPZ height distribution.}
\label{imh+kpz}
\end{center}
\end{figure}

The same kind of patterns can be also observed in case of strong
KPZ anisotropy $p=1$, $q=0$ for short times, but later the dots are
smeared. For the LHOD+KPZ case a very slow crossover to 
KPZ scaling (see Fig.~\ref{imh+kpz}) occurs.
Although the data collapse with KPZ exponents is rather poor for
smaller sizes (it would be better with larger $z$ and $\alpha$ exponents
corresponding to the MBE class), for $L\ge 512$ it agrees with KPZ.
The wavelength saturates very quickly (see insert of Fig.~\ref{imh+kpz})
in agreement with KPZ, where no coarsening is expected.

\begin{figure}[ht]
\begin{center}
\epsfxsize70mm
\epsffile{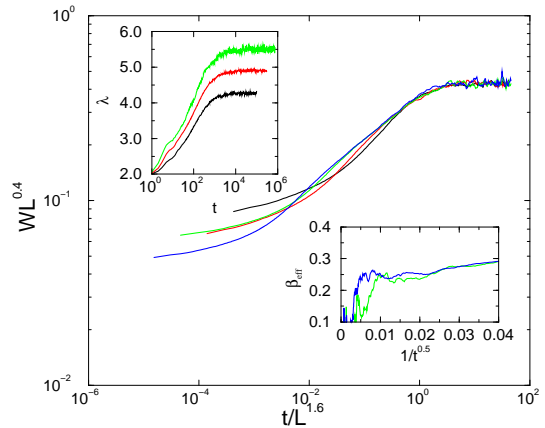}
\caption{(Color online) Data collapse of the $L=128,...1024$ 
(top to bottom at the right) LCOD model ($p_{\pm x}=p_{\pm y}=1$)
with competing deposition ($p=1$). One can see clear KPZ scaling. 
The left insert shows the logarithmic growth of $\lambda$ for 
$L=128,256,512$ (bottom to top). Right insert: $\beta_{eff}$ as
the function of time.}
\label{KSB}
\end{center}
\end{figure}

Having confirmed that the LCOD model exhibits MH scaling,
now we can investigate the scaling behavior of the (inverse) 
KS equation (\ref{KS-e}), described by the combination 
of inverse MH and normal KPZ processes. We have run extensive 
simulations up to $t=3\times 10^6$ MCs (for $L=128,256,512,1024$) 
to obtain firm numerical evidence.
As Fig.~\ref{KSB} shows the finite size scaling collapse with 
with KPZ exponents is satisfied and the effective $\beta$ extrapolates
to $1/4$. This value agrees well with our high precision KPZ simulation
result \cite{asepddcikk}.
The wavelength grows logarithmically in time (\ref{lamloggrow})
(see insert of Fig.~\ref{KSB}) and saturates well before the
steady state. In the steady state it grows slowly with the system 
size as (\ref{lamlogrou}).
In a one decade long time window one can fit the data with 
$\lambda(t,L\to\infty) \propto t^{0.12(2)}$, but due to the clear 
logarithmic behavior in the steady state (\ref{lamlogrou}), 
one should not take such power-law fitting very seriously.

For the sake of completeness we have performed similar analysis 
for the anisotropic LCOD model ($p_{\pm x}=1, p_{\pm y}=0$) with KPZ
too, for sizes $L\le 1024$. We have obtained agreement with the 
KPZ scaling for the width $W$ and logarithmic growth for $\lambda$ as
in case of isotropy. 

\section{KPZ in the presence of normal surface diffusion}

\begin{figure}[ht]
\begin{center}
\epsfxsize70mm
\epsffile{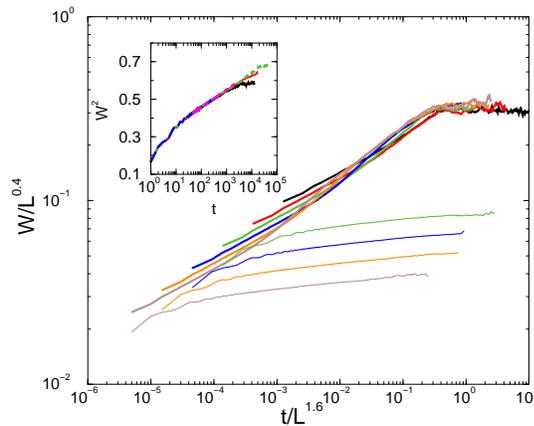}
\caption{(Color online) Data collapse of KPZ deposition ($p=1$) and
weak, isotropic normal LHOD (higher curves) for $L=64,128,...2048$ 
(top to bottom). In case of strong diffusion (lower curves) 
the KPZ scaling disappears and as the insert shows logarithmic
growth can be observed. }
\label{mh-kpzw}
\end{center}
\end{figure}

In this part we show the scaling behavior study of the KPZ process 
in the presence of normal (smoothing) LHOD, introduced in the previous 
section. First let's consider the weak isotropic diffusion case: 
$p_{\pm x}=p_{\pm y}=0.1$.
As Fig.~\ref{mh-kpzw} shows the $W(t)$ curves for sizes
$L=64,128,...2048$ exhibit a good collapse when we rescale them 
with the $2+1$ dimensional KPZ exponents (\ref{2dkpzexps}). 
This means that the KPZ scaling is stable against of the introduction
a smoothing, MBE type of surface diffusion.

On the other hand when we add strong LHOD diffusion, 
$p_{\pm x}=p_{\pm y}=0.9$, to the KPZ process ($p=1$) the surface 
growth slows down and we don't find the KPZ scaling any more
(see the lower part of Fig.~\ref{mh-kpzw}). 
Instead, logarithmic surface growth emerges, as shown in the insert 
of Fig.~\ref{mh-kpzw}. A fitting with the form $W^2(t)=A+B\ln(t)$ 
for the asymptotic growth regime gives $A=0.40(1)$ and $B=0.26(1)$. 
The amplitude of this growth is different from the 
exactly known universal value of the EW class in two dimensions:
$A_{EW} = 0.151981$ \cite{NatTang92}.

\begin{figure}[ht]
\begin{center}
\epsfxsize70mm
\epsffile{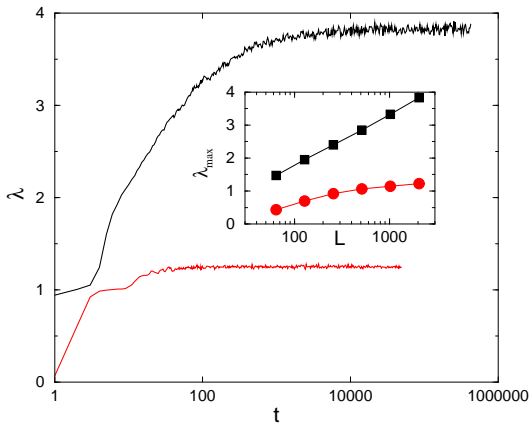}
\caption{(Color online) The wavelength saturates quickly for KPZ + weak 
LHOD (higher curve) and KPZ + strong LHOD (lower curve) diffusion ($L=2048$).
The insert shows $\lambda_{max}$ versus $L$.}
\label{KSl8}
\end{center}
\end{figure}

The wavelength saturates very quickly to the maximal value, which for 
weak diffusion scales logarithmically with the system size  
(\ref{lamlogrou}), while in case of the strong diffusion the
characteristic length remains on the order of lattice unit:
$\lambda_{max} \simeq 1$, with a very weak size dependence, corresponding
to uncorrelated surface heights (see Fig.~\ref{KSl8}).
It is well known, that in the strong diffusion limit, the relevant 
fluctuations below $d_c$ can be washed away, hence the logarithmic 
surface growth we observe should correspond to the mean-field 
behavior of the strong-coupled KPZ. This provides us a unique way
to study the crossover behavior between the KPZ and KPZ mean-field
behavior.

\section{Probability distribution results }

Here we present the probability distributions $P(W^2)$ obtained in the
steady state of our model. Such distributions are universal, hence they 
complement the previous scaling results. The exact functional form for 
KPZ is known in one dimension only, but in two dimensions very precise 
numerical data exist, obtained via other surface models \cite{RP94}. 
The distributions of those KPZ models have been determined in higher 
dimensions \cite{MPPR02}, suggesting the lack of finite upper 
critical dimension.

\begin{figure}[ht]
\begin{center}
\epsfxsize70mm
\epsffile{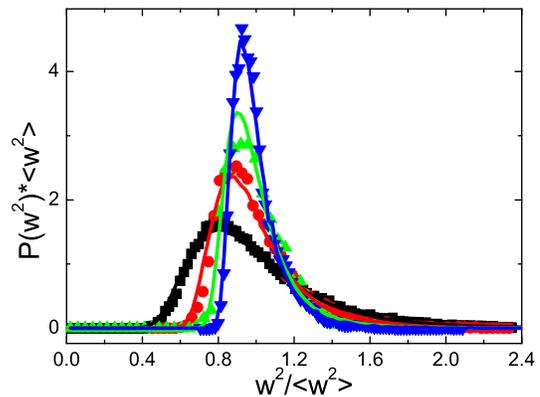}
\caption{(Color online) Comparison of the $P(W^2)$ of the higher dimensional
octahedron model results (symbols) with those of \cite{MPPR02} (lines) in 
$d=2,3,4,5$ spatial dimensions (bottom to top).}
\label{zoltan_cmp}
\end{center}
\end{figure}

First, we compare our $P(W^2)$ results for KPZ with those of \cite{MPPR02} 
in $d=2,3,4,5$ dimensions. The $W^2$ distribution data were taken from 
the saturation regimes and analyzed in systems of sizes: $L=1024$ (2d), 
$L=512$ (3d), $L=64$ (4d), $L=32$ (5d).
The presented data are coming from our higher dimensional KPZ simulations,
using the extended octahedron model described in \cite{asepddcikk}.
When we rescaled our data with $\langle W^2\rangle$ as Fig.~\ref{zoltan_cmp}
shows we found very good agreement in $d=2,3,4,5$ dimensions with the
earlier KPZ distribution curves.
Again, we can't see a signal for an upper crictical dimension at $d_c=4$,
conjectued by theoretical approaches (see for example \cite{F05}).

\begin{figure}[ht]
\begin{center}
\epsfxsize70mm
\epsffile{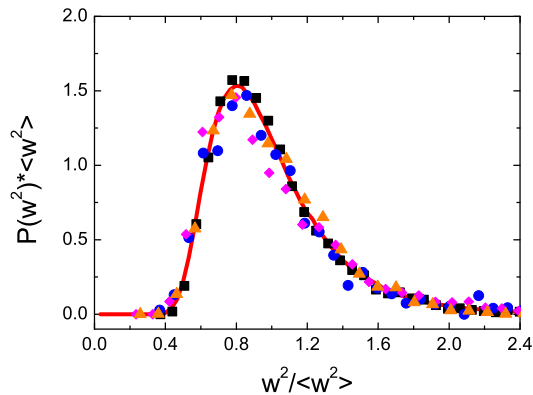}
\caption{(Color online) Comparison of $P(W^2)$ of the KPZ+LHOD 
(black boxes); KPZ+inverse LHOD (blue dots); 
KPZ+inverse, anisotropic LHOD (pink rhombuses); 
KPZ+inverse LCOD (orange triangles) with that of the 
KPZ from ref.\cite{MPPR02} (solid line).}
\label{KShist}
\end{center}
\end{figure}

Furthermore, we tested our surface scaling results for KPZ with
in the presence of diffusion.
In case of competing KPZ and LHOD/LCOD processes (i.e. for $p=1$, $q=0$, 
$p_{\pm x}=p_{\pm y}=1$) we determined the $P(W^2)$ distributions 
well in the saturation regime of systems of size $L=1024$. The steady state
could be reached for: $t> \sim 10^5$ MCs in case of KPZ+LHOD and for 
$t>5\times 10^7$ MCs in case of the KPZ+LCOD model. We generated $100$ 
independent samples, and cut out the steady state data of $W^2(t)$.
We calculated the scaled steady state probability distributions as
shown on Fig.~\ref{KShist} for different combinations.
The comparison with the histogram of the $2+1$ dimensional KPZ from
\cite{MPPR02} shows a good agreement in general, providing further
numerical evidence that the KS model asymptotically exhibits the 
KPZ universality class scaling. Our results complement the one-dimensional 
simulations results of \cite{krug-hont}, in which the equivalence of 
KS and KPZ scaling was confirmed numerically. 

\section{Conclusions and outlook}

We have returned to some unresolved questions of basic surface growth
phenomena using extensive computer simulations of atomistic models.
Contrary to earlier studies we could perform numerical analysis of
models with surface diffusion in $2+1$ dimensions, due to the effective
mapping of RSOS models onto binary lattice gases.

We have shown that in the zero external noise limit RSOS models 
with short range interactions can be constructed, which exhibit
Molecular Beam Epitaxy or Mullins Herring type of surface growth.
For inverse (roughening) diffusion, which increases the local curvature,
unstable growth, resulting in pyramid-like structures emerges. The size 
of these structures is limited only by $L$, which is not directly 
comparable with real materials. We created these MBE or MH type of 
atomistic models in order to study them in a competition with
non-conserved KPZ processes. The simulations provided numerical
evidence that strong, smoothing surface diffusion can slow down the 
KPZ to a logarithmic growth, thus we are able to reach the mean-field 
behavior of the strongly coupled KPZ fixed point in two dimensions, 
which is expected to show up in high dimensions only. 

The mapping of the surface models onto lattice gases implies that the 
(anisotropic), oriented diffusion of dimers (KPZ) is stable against 
the introduction of an attracting force among them, but a
strong repulsion can destroy the fluctuations, resulting in a
mean-field behavior.
We provided strong numerical evidence using surface scaling and
probability distribution studies, that the KS model exhibits KPZ 
scaling in $2+1$ dimensions as conjectured by field theory.
We summarized the models we considered in case of spatially
isotropic surface diffusion and non-conserved noise in Table~\ref{tab}.
 
Further studies of LCOD, with different boundary conditions can set the target
of the research of the surface inclination by mapping the surface tilt 
onto the total particle concentration of the lattice gas.
In particular the angle dependence of the phase transitions among
different growth phases can be understood by considering
the underlying driven gas.
Using our method one can transform results of disorder or of anomalous 
diffusion between the surface and lattice gas models.

We introduced a characteristic length scale $\lambda$ to follow the 
dynamics of patterns, which occur, if normal (smoothing) KPZ competes 
with inverse (roughening) diffusion. We investigated this 
race for MH and MBE process, with and without spatial anisotropies. 
In case of uni-axial surface diffusion ripple, while for $x/y$ lattice 
isotropy, dot like pattern formation could be achieved. The wavelength 
growth is slow and saturates much before the steady state. 
Usually, we found logarithmic time and system size dependence of $\lambda$, 
except when steady DC current flows through the system. In this case
the interfaces are more rough, the ripples are bended and power-law 
scaling is observable. In this case the scaling of $\lambda$ agrees with the 
scaling of the width, i.e. characterized by the MH class exponents.
This finding agrees with 3d kinetic Monte Carlo simulations of epitaxial 
growth and erosion on (110) crystal surfaces \cite{GL02} and with
analytic arguments \cite{GL08}. Furthermore in case of ion beams
with grazing incidence the dislocation dynamics results in such
athermal, kinetic coarsening of the patterns \cite{H09}.

The wavelength behavior can be understood with the help of considering
the underlying lattice gas model, since the ripple or dot structures
correspond to a phase separation. The coalescing surface pattern dynamics 
can be mapped onto the generalization of the reaction-diffusion process
\cite{Obook08} of extended objects. It was shown that in ASEP type of models,
where strong phase separation is present the domain growth follows slow, 
logarithmic behavior \cite{EKKM98,KBEM00} in case of smooth surfaces.
On the other hand for rough surfaces, power-law coarsening of $\lambda$ 
has been derived using simple scaling arguments \cite{EKLD02}.

Finally we mention that our models enable effective, bit-coded, stochastic 
cellular automaton type of simulation of surfaces, hence they could be
run extremely fast on advanced, graphic cards. 

\begin{table*}
\begin{ruledtabular}
\caption{Overview of the surface models with spatial isotropy and non-conserved 
noise. Columns 2-5 show the sign of couplings of the differential equations considered.}
\begin{tabular}{|c|c|c|c|c|c|}
model acronym & $\sigma$ & $\lambda_2$ & $\lambda_{22}$ & $\nu_4$ & properties\\ \hline
EW    &  +       &       0     &     0          &    0    & smooth surface (logarithmic) growth \\
KPZ   &  +       &      +/-    &     0          &    0    & rough surface (power-law) growth \\
MH    &  0       &       0     &     0          &    +    & smoothing diffusion, $z=4$ \\
iMH   &  0       &       0     &     0          &    -    & inverse/roughening diffusion, unstable, $z=4$ \\
MBE   &  0       &       0     &     +/-        &    +    & normal/smoothing diffusion, $z=10/3$ \\
iMBE  &  0       &       0     &     +/-        &    -    & inverse/roughening diffusion, $z=10/3$, pyramids\\
KS    &  +       &      +/-    &     0          &    -    &  KPZ scaling, dot pattern (logarithmic) growth \\
iKS   &  +       &      +/-    &     0          &    +    &  KPZ scaling, mean-field for strong diffusion \\
\hline
\end{tabular}\label{tab}
\end{ruledtabular}
\end{table*}

\vskip 1.0cm

\noindent
{\bf Acknowledgments:}\\

We thank Zolt\'an R\'acz for the useful comments and providing us
universal probability distribution functions $P(W)$ of other KPZ models.
Support from the Hungarian research fund OTKA (Grant No. T046129),
the bilateral German-Hungarian exchange program DAAD-M\"OB 
(Grant Nos. D/07/00302, 37-3/2008) is acknowledged. 
The authors thank for the access to the HUNGRID, Cluster-grid 
and the NIIF supercomputer in Budapest.

\bibliography{ws-book9x6}

\end{document}